\documentclass[pageno]{sig-alternate-05-2015}

\usepackage{mathptmx}
\newcommand{\ignore}[1]{}
\usepackage{fancyhdr}
\usepackage{amssymb,amsmath,amsfonts}
\usepackage{hyperref}
\usepackage{color}
\usepackage[table]{xcolor}
\usepackage{paralist}
\usepackage{comment}
\usepackage{listings}
\usepackage{xspace}
\usepackage{multirow}
\usepackage[capitalize]{cleveref}
\usepackage{graphicx}
\usepackage{subfig}
\usepackage{tikz}
\usepackage[linesnumbered, ruled]{algorithm2e}
\usepackage{algorithmic}
\usepackage{ulem}

\newcolumntype{P}[1]{>{\centering\arraybackslash}p{#1}}

\newcommand{\eg}{e.g.}
\newcommand{\ie}{i.e.}

\newcommand{\pt}{\textit{pt}\xspace}
\newcommand{\ts}{\textit{ts}\xspace}
\newcommand{\pts}{\textit{pts}\xspace}
\newcommand{\wts}{\textit{wts}\xspace}
\newcommand{\wpt}{\textit{wpt}\xspace}
\newcommand{\rts}{\textit{rts}\xspace}

\newcommand{\lts}{\textit{lts}\xspace}
\newcommand{\sts}{\textit{sts}\xspace}

\newcommand{\acquirets}{\textit{acquire\_ts}\xspace}
\newcommand{\releasets}{\textit{release\_ts}\xspace}
\newcommand{\maxts}{\textit{max\_ts}\xspace}
\newcommand{\lease}{\textit{lease}\xspace}

\newtheorem{definition}{Definition}
\newtheorem{theorem}{Theorem}
\newtheorem{lemma}{Lemma}
\newtheorem{assumption}{Assumption}
\newcommand{\M}{\textit{M}\xspace}
\newcommand{\Sh}{\textit{S}\xspace}

\title{Tardis 2.0: Optimized Time Traveling Coherence for Relaxed Consistency Models}
\author{Xiangyao Yu$^1$, Hongzhe Liu$^2$, Ethan Zou$^3$, Srinivas 
 Devadas$^1$
  \\
    $^1$ Massachusetts Institute of Technology  \\
    $^2$ Algonquin Regional High School \ \ \ \  $^3$ Lexington High School 
\\
$\{$yxy, devadas$\}$@mit.edu, $\{$liuhenry4428, 
zouethan20$\}$@gmail.com }

\begin{document}
\CopyrightYear{2016}
\setcopyright{acmcopyright}
\conferenceinfo{PACT '16,}{September 11-15, 2016, Haifa, Israel}
\isbn{978-1-4503-4121-9/16/09}\acmPrice{\$15.00}
\doi{http://dx.doi.org/10.1145/2967938.2967942}

\date{}
\maketitle
\thispagestyle{empty}

\begin{abstract}

Cache coherence scalability is a big challenge in shared memory 
systems. Traditional protocols do not scale due to the storage and 
traffic overhead of cache invalidation.  Tardis, a recently proposed 
coherence protocol, removes cache invalidation using logical 
timestamps and achieves excellent scalability.
The original Tardis protocol, however, only supports the Sequential 
Consistency (SC) memory model, limiting its applicability.
Tardis also incurs extra network traffic on some benchmarks due to 
renew messages, and has
suboptimal performance when the program uses spinning to 
communicate between threads.

In this paper, we address these downsides of Tardis protocol and make 
it significantly more practical.
Specifically, we discuss the architectural, memory system and protocol changes
required in order to implement the TSO consistency model on Tardis, and 
prove that the modified
protocol satisfies TSO.  We also describe modifications for Partial Store Order (PSO) and Release Consistency (RC). Finally, we propose optimizations for better 
leasing policies and to handle program spinning.
On a set of benchmarks, optimized Tardis improves on
a full-map directory protocol in the metrics of performance,
storage and network traffic, while being simpler to implement.
\end{abstract}

\section{Introduction} \label{sec:intro}

As the number of cores on a single chip increases,
the cache coherence protocol becomes a potential scalability and 
performance bottleneck. \textit{Snoopy} coherence 
protocols~\cite{snoopy} work well for small-scale systems with a few 
cores, but do not scale due to the
traffic pressure caused by broadcasting messages on the bus.  
\textit{Directory-based} coherence protocols~\cite{censier1978, 
tang1976} have better scalability and are widely used in multicore 
processors today~\cite{tilera, xeonphi}.  For future systems with 
hundreds or even thousands of cores, however, the storage overhead of 
a full-map directory becomes a serious scalability bottleneck.  A 
number of enhancements to directory coherence protocols have been 
proposed in the literature~\cite{kelm2010, sanchez2012, ATAC, maa1991} 
to improve scalability. These enhancements typically sacrifice performance and 
incur extra implementation and verification complexity. 

As an alternative, a recently proposed cache coherence protocol called 
\textit{Tardis} shows better scalability while maintaining simplicity 
and high performance~\cite{tardis}. The key insight behind Tardis is 
that it is sufficient but \textit{unnecessary} to enforce the global 
memory order 
in physical time order, which is what traditional coherence protocols 
do. Instead, Tardis uses a combination of logical timestamps and
physical time. In Tardis, a memory write does not invalidate shared 
copies in other cores' private caches. Instead, the write is 
immediately performed at a logical timestamp greater than the 
timestamps of all the shared copies. The new data version can coexist 
with old versions at the same physical time but is ordered after the 
old versions in logical time.


Introducing logical timestamps into the coherence protocol brings 
several salient properties to Tardis. First, Tardis is 
\textit{scalable}; it has no sharer list and only requires 
$O(\log{N})$ storage per cacheline for an $N$-core system. It also 
does not require
broadcasting support. 
Tardis is also \textit{simple}; it uses timestamps to explicitly 
express the consistency model and is easy to reason about and 
verify~\cite{tardis-proof}. Different from other protocols, 
Tardis does not need invalidation and is therefore able to achieve 
better performance on some benchmarks.

There are three drawbacks of the original Tardis protocol. The first 
drawback is that it only supports the \textit{sequential consistency} (SC) 
memory model. While SC is simple and well studied, commercial 
processors usually implement more relaxed models. Intel 
x86~\cite{sewell2010} and SPARC~\cite{weaver1994} processors can 
support \textit{Total Store Order} (TSO); ARM~\cite{seal2001} and IBM 
Power~\cite{mador2012} processors implement weaker consistency models.  
It is difficult for
commercial processors to adopt Tardis if the target memory models 
are not supported.

Another drawback of Tardis is the renew message that is used to
extend the logical lease of a shared cacheline in a private cache.  
These messages incur extra latency and bandwidth overhead. 
Finally, Tardis uses a timestamp self-increment strategy to avoid livelock.
This strategy has suboptimal performance when threads communicate via
spinning.

In this paper, we will address these drawbacks of the original Tardis 
protocol and make it more practical. Specifically, we will discuss the 
changes to the cores and the memory subsystem in order to implement 
TSO and other relaxed consistency models on Tardis. A formal proof 
that our algorithm correctly implements TSO is also given. We also 
propose new optimization techniques (MESI, livelock detection and 
lease prediction) to reduce the number of renew messages in Tardis, 
which improves performance. 

Our simulations over a wide range of benchmarks indicate that our 
optimizations improve the performance and reduce the network traffic 
of Tardis. Compared to a full-map directory-based coherence protocol 
at 64 cores, optimized Tardis is able to achieve better 
performance (1.1\% average improvement, up to 9.8\%) and lower network 
traffic (2.9\% average reduction, up to 12.4\%) at the same time. The 
improvement increases as the system scales to 256 cores where Tardis 
improves performance (average 4.7\%, upto 36.8\%) and reduces the network 
traffic (average 2.6\%, upto 14.6\%).  While the optimized and baseline 
Tardis protocols require timestamps in the L1 cache, they are, 
overall, more space-efficient than full-map directory protocols and 
simpler to implement.

\section{Background} \label{sec:background}

In this section, we introduce the key concepts behind
the basic Tardis protocol (\cref{sec:physiological} and \cref{sec:sc}) and
the protocol itself (\cref{sec:tardis-sc} and \cref{sec:example-sc}). We 
describe the pros and cons of Tardis vis-\`{a}-vis a directory 
coherence protocol (\cref{sec:tardis-vs-dir}).

\subsection{Physiological Time} \label{sec:physiological}

Most consistency models define a global order for all memory loads and 
stores. Traditional cache coherence protocols enforce this global 
memory order using physical time order. This requires an invalidation 
mechanism for write-after-read data dependency.  Specifically, a write 
to a shared cacheline has to invalidate all the shared copies of that 
line because the write must be ordered after all previous reads with 
respect to physical time. The invalidation mechanism requires either 
broadcasting support in the network (as in snoopy coherence protocols) 
or a per cacheline sharer list in the directory (as in directory-based 
coherence protocols) and is therefore not able to scale to large 
numbers of cores. 

Tardis avoids invalidation by introducing logical timestamps
into the cache coherence protocol and enforcing the global memory 
order in both logical and physical time order.  In Tardis, a write 
does not invalidate shared copies, instead, a new version of the 
cacheline is created at a later logical time at which point all 
previous shared copies have expired.  This usage of logical time to 
order data bears some similarity to Lamport clock~\cite{lamport1978} 
and the timestamp ordering~\cite{bernstein79, reed78} concept in 
database concurrency control algorithms.

Specifically, Tardis invented a new concept of time called 
\textit{physiological time} and enforces the global memory order with 
respect to it. Using $<_{pt}$ and $<_{ts}$ to represent physical time 
and logical timestamp order respectively, the physiological time order 
($<_{pl}$) is defined by \cref{eq:pl}.

\vspace{-0.2in}
\begin{equation} \label{eq:pl}
X <_{pl} Y := X <_{ts} Y\ or\ (X =_{ts} Y\ and\ X <_{pt} Y) 
\end{equation}
\vspace{-0.2in}

An operation $X$ is before $Y$ in physiological time order if $X$ has 
a smaller timestamp or if they have the same timestamp but $X$ happens 
earlier in physical time. If two operations happen at the same logical 
and physical time, then we can use any metric to break the tie (since 
they do not have a conflict). Physiological time makes it 
easy to reason about consistency models implemented on Tardis.

%
%

\subsection{Sequential Consistency (SC)} \label{sec:sc}

The original Tardis protocol~\cite{tardis} implements sequential 
consistency, which requires that {\it ``the result of any execution is 
the same as if the operations of all processors (cores) were executed 
in some sequential order, and the operations of each individual 
processor (core) appear in this sequence in the order specified by its 
program''}~\cite{lamport1979}.  

Sequential consistency can also be defined using the following two 
axiomatic rules, where $X$ and $Y$ are two memory operations, $<_p$ 
and $<_m$ are the program and global memory order respectively, and 
$L(a)$ and $S(a)$ are load and store to address $a$ respectively.

\textbf{Rule 1}: $X <_p Y \Rightarrow X <_m Y $

\textbf{Rule 2}: Value of $L(a) = $ Value of Max$_{<m}$ \{$S(a)$ | 
$S(a) <_m L(a)$\}

Rule 1 says that the sequential order should agree with the program 
order and Rule 2 says that each load should observe the preceding  
store in the sequential order. 

In Tardis, the global memory order ($<_m$) in Rule 1 of SC can be 
implemented using the physiological time order ($<_{pl}$).  
\cref{sec:tardis-sc} describes in detail how Tardis implements SC. 

\subsection{Tardis with SC} \label{sec:tardis-sc}

This section presents the baseline Tardis protocol proposed in 
\cite{tardis}. For simplicity, we assume a multicore architecture with 
a private L1 cache per core and a shared L2 as the Last Level Cache 
(LLC).  We also assume a single-threaded in-order core model where an 
instruction can be issued only after the previous instruction has 
finished its execution and committed. In practice, Tardis can be 
implemented on multi-threaded out-of-order cores as well.  The 
physical time when an instruction commits is its \textit{physical 
commit time}. Each instruction also has a \textit{logical commit 
timestamp} indicating the logical time when it commits.  For the rest 
of the paper, the term ``timestamp'' always indicates logical time.  
Also, when we say an operation is performed at a certain 
physical/logical time, we mean the physical/logical commit time of 
that operation.

\subsubsection{Timestamp Management}

Although both physical and logical commit time are used for reasoning 
about Tardis. Physical commit time is not explicitly stored in 
hardware and thus there is no need to read from CPU clocks. This is an 
important property which makes Tardis scalable. 
In Tardis, physical time order is meaningful only for operations that 
already happen in physical time order, e.g., instructions committed 
from the same core, or a load observing a preceding store. Orders 
between other operations are enforced using logical timestamps.  

Logical timestamps in Tardis are explicitly stored and managed in 
hardware. For SC, each thread maintains a program timestamp (\pts) as 
the logical commit timestamp of the last operation in the program 
order. For multithreaded cores, each hardware context maintains its 
own \pts in hardware.  According to Rule 1 in the definition of SC, 
the global memory order (i.e., physiological commit time order in 
Tardis) must agree with the program order.  Therefore, the \pts should 
monotonically increase with the program order\footnote{So for every $X 
<_p Y$, we have $X \leq_{ts} Y$ and $X <_{pt} Y$ (due to in-order 
processing) and therefore $X <_{pl} Y$.}. 

Each cacheline in Tardis maintains a write timestamp (\wts) and a read 
timestamp (\rts). \wts is the commit timestamp of the last store and 
\rts can be considered the commit timestamp of the last load.  
Together they mean that the value is valid between \wts and \rts in 
logical time.  Without further optimizations, for a cache with (\wts, 
\rts), a load can only be performed at \ts
such that $\wts \leq \ts \leq \rts$, and a store can only be performed 
at \ts such that $\ts \geq \rts + 1$. In terms of logical time, this 
means that a load always observes the preceding store and only one 
version of data exists at any timestamp. 

\subsubsection{Tardis-SC Protocol} 

We now present the baseline Tardis-SC protocol. For clarity, we first 
explain how shared caching is implemented in Tardis-SC and then 
discuss exclusive caching.    

\begin{figure*}[ht!]
	\centering
	\includegraphics[width=.95\textwidth]{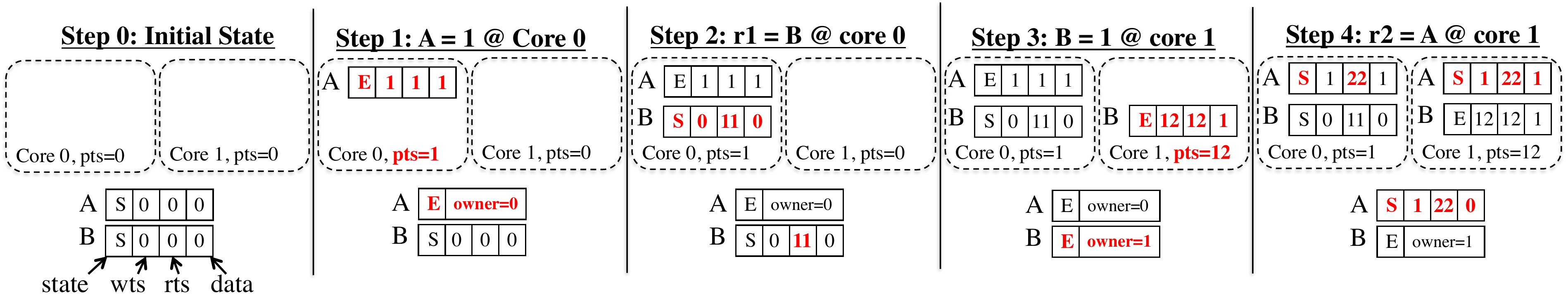}
	\vspace{-.04in}
	\caption{ Tardis-SC example (\lease$ = 10$).  Cachelines in 
private caches and LLC are shown. Changed states at each step are 
in bold red.}
	\vspace{-.2in}
\label{fig:example-sc}
\end{figure*}

\textbf{Shared Caching}

An important concept in Tardis is \textit{timestamp leasing} of a 
cacheline. In order to load a shared copy into a core's private cache, 
the core needs to get a lease on the cacheline. Multiple cores can 
cache the same cacheline in shared state with overlapping leases. But 
an update can be performed on the cacheline only after all shared 
leases expire.  Leasing in Tardis is implemented with respect to 
logical but not physical time.  

In private caches, the \rts of a cacheline indicates the end of the 
lease on the line. A load operation can be performed as long as $\pts 
\leq \rts$ and the load updates \pts to \textit{Max(\pts, \wts)}. If 
$\pts > \rts$, however, the load cannot be directly performed (for 
example, the line may have been modified in the LLC after \rts but 
before \pts in logical time) and a renew request has to be sent to the 
LLC to extend the \rts of the data version. If the version in the LLC 
is the same as the version in the L1 (i.e., they have the same \wts), 
the extension is successful and the extended \rts is returned.  
Otherwise, the extension fails and the latest version is returned 
instead. But regardless, the returned \rts will be greater than \pts 
and the load can now be performed. The baseline Tardis uses 
speculative loading to hide the latency of renew messages.  

In the shared LLC, the \rts of a cacheline is the maximal \rts among 
all the shared copies of the same address, namely, the end of all 
shared leases. A load or renew request from a core to the LLC may 
extend the cacheline's \rts if the requesting core has a sufficiently 
large \pts.  Specifically, the \rts is updated to \textit{Max(\rts, 
\pts + \lease)} for the load request where \lease is a constant (e.g., 
\lease = 10) in
the baseline Tardis. The \lease is chosen to be non-zero so that the 
line will be valid in the L1 for longer logical time and therefore not 
be constantly renewed by the core. The data and both \wts and \rts are 
returned to the requesting core in the response.  

In Tardis, if a core keeps reading a stale cacheline at a \pts smaller 
than the cacheline's \rts, the core may never observe a later update 
and livelock may occur. To guarantee that a write is eventually 
observed by other cores, Tardis forces the \pts of each core to 
periodically increase so that a stale version of a cacheline will 
eventually expire. We return to the subject of livelock in 
\cref{sec:livelock}.

\textbf{Exclusive Caching}

Exclusive caching in Tardis is similar to the directory baseline for 
the most part. For example, a write operation needs to acquire 
exclusive ownership from the LLC; ownership transfers if a another 
core writes to the same cacheline; if another core reads the 
cacheline, the owner downgrades the line to shared state and writes 
the data back to the LLC.  For a write operation at an L1, the newly 
created version has \textit{new\_version.\wts = previous\_version.\rts 
+ 1}.    

When a core writes to a cacheline that is shared by one or
more cores, however, Tardis has fundamentally different behavior than 
a traditional cache coherence protocol. When the write request reaches 
a shared cacheline in the LLC, unlike traditional cache coherence 
protocols, Tardis does not send invalidations to the sharers. Instead, 
exclusive ownership can be immediately returned to the requesting 
core. The core can perform the write at a \wts greater than the 
cacheline's current \rts. Since all the shared copies expire at this 
\rts, this current write operation happens at a timestamp greater than 
all reads to the previous version. Therefore, the write is after those 
reads in the global memory order. Given the definition of SC 
(Rule 2 in particular), this behavior is sequentially consistent 
according to physiological time order.    

\subsection{Tardis-SC Example} \label{sec:example-sc}

\vspace{-.2in}
\begin{lstlisting}[language=C,label={lst:example-sc},caption={Example 
Program}]
           initially A = B = 0
        [Core 0]          [Core 1]
         A = 1             B = 1
         r1 = B            r2 = A
\end{lstlisting}

\cref{fig:example-sc} shows how the dual-thread program in 
Listing~\ref{lst:example-sc} is executed in Tardis-SC. For simplicity, 
the example executes the four operations one at a time, and executes 
the operations from core 0 before those from core 1. Initially, both 
cores' L1s are empty; all cachelines in the LLC are in shared state;
and all timestamps are 0. 

\textbf{Step 1:} Core 0 stores to address $A$. A store request is sent 
to the LLC. Since A is in shared state in the LLC, exclusive ownership 
is immediately returned to core 0.  Core 0 performs the store at 
timestamp $A.\rts + 1 = 1$.  After the store, core 0's \pts becomes 1.  
The new version of $A$ has $\wts = \rts = 1$.

\textbf{Step 2:} Core 0 loads from address $B$. A load request is sent 
to the LLC with $\pts = 1$. $B$'s \rts in the LLC is extended to 
$Max(\rts, \pts + \lease) = 11$. The data value and both timestamps 
are returned to Core 0's L1 cache.   

\textbf{Step 3:} Core 1 stores to address $B$. A store request is sent 
to the LLC. Since $B$ is in shared state, exclusive ownership is 
immediately returned to core 1, which performs the store at timestamp 
$A.\rts + 1 = 12$ and updates the \pts, \wts and \rts. Note that 
although Core 0 is caching the previous version of $B$, no 
invalidation is sent for Core 1's store. After the store is performed, 
there are actually two different versions of $B$ coexisting in the 
core 0's and core 1's L1 caches. However, they are valid at different 
logical timestamps. Core 0's copy is valid from timestamp 0 to 11, 
while core 1's copy is valid from timestamp 12. Therefore, sequential
consistency is still maintained in physiological time order.  

\textbf{Step 4:} Core 1 loads from address $A$. A load request is sent 
to the LLC. Since Core 0 is the current owner, a message is sent to 
Core 0 asking for data and \wts/\rts writeback. The LLC is updated 
based on the response and a shared copy is returned to Core 1. Since 
the \pts of the load request is $12$, the \rts of all $A$'s cachelines 
are extended to $\pts + \lease = 22$.

\subsection{Tardis vs. Directory-based Coherence}
\label{sec:tardis-vs-dir}

A key advantage of Tardis compared to a traditional physical time 
based coherence protocol is the removal of the invalidation mechanism.   
This simplifies protocol design, reduces storage overhead, and can 
potentially achieve better performance.  

Removing the invalidation mechanism means that the LLC does not notify 
the sharing cores when a cacheline is modified. So each core should 
contact the LLC to learn the freshness of a cacheline.  This is done 
through the renew messages (cf. \cref{sec:tardis-sc}).  In a sense,
replacing invalidations with renewals is replacing a push-based model 
with a poll-based model. A poll-based model is usually simpler and  
more scalable, but may lead to wasted messages. For example, read only 
data may be constantly renewed due to expiration, but these messages 
would not exist in an invalidation-based protocol. The latency and 
bandwidth overhead of the extra renew traffic is the major potential  
disadvantage of Tardis compared to directory coherence protocols. 

Some optimization techniques (\eg, speculative read) have been 
proposed to hide the latency of renewals~\cite{tardis}.  However,
the performance and network traffic overhead can still be significant 
for some benchmarks. We will present our solutions to the renewal
problem in \cref{sec:renew}. 

\section{Tardis with TSO} \label{sec:tso}

The original Tardis protocol only supports the sequential consistency 
(SC) memory model. Although SC is intuitive and precisely defined, it 
may overly constrain the ordering of memory operations. In practice, 
this may lead to suboptimal performance. To resolve this disadvantage 
of SC, relaxed consistency models have been proposed and widely 
implemented in real systems. Most of these models focus on the 
relaxation of the program order in SC (Rule 1 in the SC definition).  
Specifically, the program order of a core may appear out-of-order in 
the global memory order. The more relaxed a model is, the more 
flexibility it has to reorder operations, which usually leads to 
better overall performance. 

In this section, we show how Tardis can be generalized to relaxed 
consistency models.  We first use Total Store Order (TSO) as a case 
study since it has a precise definition and is the most widely 
adopted. We will present the formal definition of TSO 
(\cref{sec:tso-def}), the Tardis-TSO protocol (\cref{sec:tso-impl}),  
an example program (\cref{sec:example-tso}) and optimizations 
(\cref{sec:tso-opt}). Finally, we generalize the discussion to other 
memory models (\cref{sec:other-models}).

\subsection{Formal Definition of TSO} \label{sec:tso-def}

The TSO consistency model relaxed the Store $\rightarrow$ Load constraint 
in the program order. This allows a load after a store in the program 
order to be flipped in the global memory order (assuming that the load 
and the store have no data or control dependency). In an out-of-order 
processor, this means that an outstanding store does not block the 
following loads. When a store reaches the head of the \textit{Re-Order 
Buffer} (ROB), it can retire to the \textit{store buffer} and finish 
the rest of the store operation there. The loads following the store 
can therefore commit early before the store is actually done. Since 
store misses are common in real applications, this relaxation can lead 
to significant performance improvement.



Similar to SC, the definition of TSO also requires a global order 
(specified using $<_m$) of all memory instructions. However, the 
global memory order does not need to follow the program order for 
Store $\rightarrow$ Load dependency.  Specifically, TSO can be defined 
using the following three rules~\cite{sorin2011}.  The differences 
between TSO and SC are highlighted in boldface.

\textbf{Rule 1}:
L(a) $<_p$ L(b) $\Rightarrow$ L(a) $<_m$ L(b)

\hspace{0.42in}L(a) $<_p$ S(b) $\Rightarrow$ L(a) $<_m$ S(b)

\hspace{0.42in}S(a) $<_p$ S(b) $\Rightarrow$ S(a) $<_m$ S(b)

\hspace{0.42in}\sout{\textbf{S(a) $<_p$ L(b) $\Rightarrow$ S(a) $<_m$ 
L(b)}}

\textbf{Rule 2}:
Value of $L(a) = $ Value of Max$_{<m}$ \{$S(a)$ | $S(a) <_m L(a)$ 
\textbf{or S(a) $<_p$ L(a)}\}

\textbf{\textbf{Rule 3}}:
\textbf{X $<_p$ FENCE $\Rightarrow$ X $<_m$ FENCE}

\textbf{\hspace{0.42in}FENCE $<_p$ X $\Rightarrow$ FENCE $<_m$ X}

In TSO, the program order implies the global memory order only for 
Load $\rightarrow$ Load, Load $\rightarrow$ Store and Store 
$\rightarrow$ Store constraints. Since there is no Store $\rightarrow$ 
Load constraint, a load can bypass the pending store requests and 
commit earlier (Rule 1).  Although the load is after the store in the 
program order, it is before the store in the global memory order.

In an out-of-order processor, TSO can be implemented using a store 
buffer which is a FIFO for pending store requests that have retired 
from ROB.  If the address of a load exists in the store buffer, then 
the value in the store buffer is directly returned; otherwise, the 
load accesses the memory hierarchy (Rule 2).  

TSO uses a \textit{fence} instruction when a Store $\rightarrow$ Load 
order needs to be enforced (Rule 3). In a processor, a fence flushes 
the store buffer enforcing that all previous stores have finished so 
that a later committed load is ordered after stores before the fence 
in physical time. If all memory operations are also fences, then TSO 
becomes SC.  

\begin{figure*}[t!]
	\centering
	\includegraphics[width=.95\textwidth]{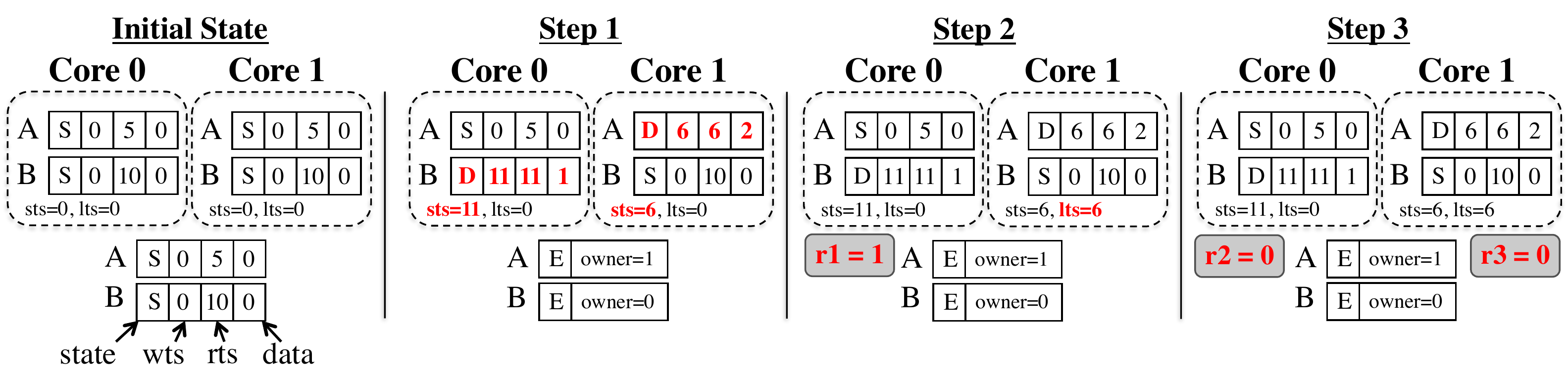}
    \vspace{-0.04in}
	\caption{ Tardis-TSO example ({\it lease} $ = 10$). A private 
cacheline in $D$ state means it is modified and dirty. Changed states 
at each step are highlighted in red.}
	\vspace{-0.07in}
\label{fig:example-tso}
\end{figure*}

\subsection{Tardis-TSO Protocol} \label{sec:tso-impl}

In this section, we describe how TSO can be implemented on Tardis.  
Specifically, we discuss the changes to the timestamp management policy 
as compared to the Tardis SC protocol. 

\subsubsection{Program Timestamp Management} \label{sec:tso-pts}

The original Tardis SC protocol uses a single program timestamp (\pts) 
to represent the commit timestamp of an instruction.  Since the 
program order always agrees with the global memory order in SC, \pts 
monotonically increases in the program order.

In TSO, however, the program order does not always agree with the 
global memory order. Following Rule 1 in TSO's definition, a store's 
timestamp is no less than the timestamps of all preceding
loads, stores and fences in the program order. A load's timestamp is 
no less than the timestamps of all preceding loads and fences, but not 
necessarily preceding stores. As a result, a single monotonically 
increasing \pts is insufficient to represent the ordering constraint. 

To express the different constraints for loads and stores 
respectively, we split the original \pts into two timestamps. The 
\textit{store timestamp} (\sts) represents the commit timestamp
of the last store, and the \textit{load timestamp} (\lts) represents 
the commit timestamp of the last load. Like \pts, both \sts and \lts 
are maintained in each core in hardware. According to Rule 1, both should 
monotonically increase in the program order because of the Load 
$\rightarrow$ Load and Store $\rightarrow$ Store constraints.  
Furthermore, the timestamp of a store (\sts) should be no less than 
the timestamp of the preceding load (\lts) because of the Load 
$\rightarrow$ Store constraint. For a load, however, its \lts can be 
smaller than \sts because there is no Store $\rightarrow$ Load 
constraint.

A fence can be simply implemented as a synchronization point between 
\sts and \lts. Specifically, a fence sets \lts $=$ \textit{Max(\lts, 
\sts)}. 
This enforces Rule 3 in TSO because operations after the fence are 
ordered after operations before the fence in physiological time order 
(and therefore the global memory order). If each memory operation is 
also a fence, then the commit timestamp for each operation 
monotonically increases and the protocol becomes Tardis SC.  


In a traditional coherence protocol, the main advantage of TSO over SC 
is the performance gain due to loads bypassing stores in the store 
buffer. In Tardis, besides bypassing, TSO can also reduce the number 
of renewals compared to SC. This is because the \lts/\sts in TSO 
may increase slower compared to the \pts in SC. As a result, fewer 
shared cachelines expire.

\subsubsection{Data Timestamp Management} \label{sec:tso-datats}


The timestamp management logic largely remains the same when the 
consistency model switches from SC to TSO. However, the timestamp 
rules for data in the store buffer need some slight changes. For 
single-threaded cores, timestamp management in the private L1 can also 
be changed for load requests for potentially better performance.  

To load a cacheline in L1 that is not \textit{dirty} (\ie, the data 
has not been changed by the core since it was cached), the timestamp 
rule is exactly the same as in SC, \ie, the \lts should fall within 
the lease of the cacheline (\wts $\leq$ \lts $\leq$ \rts); \lts jumps 
to \wts if \lts is smaller than \wts and a renewal is sent if \lts is 
greater than \rts.

For a dirty cacheline in the store buffer or L1 cache, however, the 
\lts does not have to be greater than the \wts of the cacheline. With 
respect to the global memory order, this means that the load can 
commit at an \lts smaller than the commit timestamp of the store 
creating the data (\wts). This behavior certainly violates SC but it 
is completely legal in TSO. 

According to Rule 2 of TSO, a load should return either the last store 
in global memory order or the last store in program order, depending 
on which one has a larger physiological time. Since a dirty cacheline 
was written by a store from the current core prior to the load, even 
if the load has a smaller commit timestamp than the store, Rule 2
still holds. A more formal proof of correctness is presented in 
\cref{sec:proof}.

Unlike in traditional processors, TSO can be implemented with Tardis even 
on in-order cores that do not have a store buffer. This is because 
Tardis can figure out the correct memory ordering using logical 
timestamps as will become clear in the example presented in the next section.
We note that our implementation of Tardis TSO still has a store buffer 
for better performance, but it is not required for functional
correctness. 

Note that if multiple threads can access the same private cache, then 
the above optimization for dirty L1 cachelines may not be directly 
applied in L1 cache (but it is still applied in the store buffer).  
Because a dirty line might be written by any thread sharing the L1.  
For these systems, this optimization can be turned off in the L1.

\subsection{TSO Example} \label{sec:example-tso} 


\vspace{-.1in}

\begin{lstlisting}[language=C,label={lst:example},caption={Example 
Program}, escapeinside={@}{@}]
     [core0]               [core1]
      B = 1                A = 2
      L(B) @$\rightarrow$ r1@            FENCE
      L(A) @$\rightarrow$ r2@            L(B) @$\rightarrow$ r3@
\end{lstlisting}

We use the example program in Listing~\ref{lst:example} to demonstrate 
how timestamps are managed in Tardis TSO. The execution of the program 
is shown in \cref{fig:example-tso}. For simplicity, we do not model 
the store buffer and execute one instruction per step for each core.

Initially, both addresses A and B are cached in Shared (S) state in 
both cores' private caches as well as the shared LLC. \wts of all 
lines are 0; \rts of all lines of address A are 5 and \rts of all 
lines of address B are 10.

\textbf{Step 1:} core 0 writes to address B and core 1 writes to 
address A.  Exclusive ownership of A and B are given to core 1 and 
core 0 respectively, and both stores are performed by jumping ahead in 
logical time to the end of the lease. After the stores, core 0's \sts 
jumps to timestamp 11 and core 1's \sts jumps to 6, but the \lts of 
both cores remain 0.

\textbf{Step 2:} core 0 loads address B. The value of the previous 
store from core 0 is returned ($r1 = 1$). Since B is in the dirty 
state, the load does not increase the \lts (\cref{sec:tso-datats}).  
In core 1, a fence instruction is executed which synchronizes the \lts 
and \sts to timestamp 6. 

\textbf{Step 3:} core 0 loads address A. Since its \lts is 0 which 
falls between the \wts and \rts of cacheline A, this is a cache hit 
and value 0 is returned ($r2 = 0$). In core 1, the load to address B 
also hits the L1 since its \lts = 6 falls within B's lease. As a 
result, the loaded value is also 0 ($r3 = 0$).

\cref{lst:mem-order} shows the physiological commit time for each 
instruction in \cref{lst:example}. It also shows the global memory 
order using arrows.  Physiological time is represented using 
a logical timestamp and physical time pair (\ts, \pt) where \ts is the 
commit timestamp and \pt is the physical commit time of the 
instruction.  According to the definition, ($ts_1$, $pt_1$) $<$ 
($ts_2$, $pt_2$) if $ts_1 < ts_2$ or ($ts_1 = ts_2$ and $pt_1 < 
pt_2$).  

\vspace{-0.1in}
\begin{minipage}[t]{.95\columnwidth}
\begin{lstlisting}[language=C,label={lst:mem-order},caption={Physiological 
commit time and global memory order.}, escapeinside={@}{@}]
@\vspace{-0.3in}@
     
     [core0]              [core1]
@\vspace{-0.05in}@
     (11, 1)              (6, 1)

     (0, 2)               (6, 2)

     (0, 3)               (6, 3)
@
\begin{tikzpicture}[overlay,remember picture]
\draw[->] (1.6, 0.85) to node {} (1.6, 0.55);
\draw[->] (2.5, 0.4) to node {} (5, 1.65);
\draw[->] (5.8, 1.5) to node {} (5.8, 1.2);
\draw[->] (5.8, 0.85) to node {} (5.8, 0.55);
\draw[->] (5, 0.4) to node {} (2.5, 1.65);
\end{tikzpicture}
@
\end{lstlisting}
\end{minipage}
\vspace{-0.1in}

The execution is definitely not sequentially consistent since the 
program order in core 0 is violated between \texttt{B = 1} and 
\texttt{L(B)}. But it obeys all the invariants of TSO. Note that the 
store buffer is not included in the example since we are modeling 
in-order cores, but TSO can still be implemented. This feature is not 
available in traditional physical time based coherence protocols. For 
this example, adding the store buffer will not change the hardware 
behavior. 

\subsection{TSO Optimizations} \label{sec:tso-opt}

Many optimization techniques have been proposed in the literature to 
improve performance of the basic SC/TSO consistency model. Examples 
include load speculation to hide Load $\rightarrow$ Load and Store 
$\rightarrow$ Load dependency, and store prefetch to enhance Store 
$\rightarrow$ Store and Load $\rightarrow$ Store 
performance~\cite{gharachorloo1991}. For TSO, the speculation can go 
over fences.

Tardis TSO is compatible with these optimizations. In fact, it may be 
even simpler to support them on Tardis than on traditional coherence 
protocols since the timestamps can help preserve/check memory order.  
For example, for load $\rightarrow$ load relaxation, multiple loads 
can be speculatively executed in parallel, and the \wts and \rts of 
the loaded cachelines are stored inside the core (\eg, in the ROB).  
To enforce load $\rightarrow$ load dependency, the processor only 
needs to commit instructions with ascending timestamp order (and 
reissue a load with a new timestamp if necessary).  In contrast, a 
traditional processor needs to snoop on invalidation messages in order 
to detect a speculation failure. Fence speculation can also be 
implemented in a similar way using timestamps.  In general, Tardis 
allows all memory operations to be speculatively executed arbitrarily, 
as long as their commit timestamps obey the consistency model. This 
flexibility makes it easier to reason about and implement these 
optimizations.

\subsection{Other Relaxed Consistency Models} \label{sec:other-models}

Similar to TSO, other memory consistency models 
(\cref{tab:consistency}) can also be supported in Tardis with proper 
changes to the timestamp rules.  Given the relationship between the 
program order and the global memory order, it is usually 
straightforward to adapt Tardis for different models. In this section, 
we briefly discuss Partial Store Order (PSO) and Release Consistency 
(RC) to illustrate how Tardis can adapt to them with minimal 
algorithmic change.  

\begin{table}
    \caption{ Memory order constraints for different consistency models.
    $L$ for load, $S$ for store and $F$ for fence. For release 
    consistency, \textit{rel} for release and \textit{acq} for 
    acquire.
    } \begin{center}
    { \footnotesize
        \begin{tabular}{|P{.6in}|p{.8in}|p{1.1in}|}
            \hline
        Consistency Model & Ordinary \newline Orderings & 
        Synchronization
        Orderings \\ \hline\hline

        SC & 
            L $\rightarrow$ L, L $\rightarrow$ S, \newline
            S $\rightarrow$ L, S $\rightarrow$ S & \\\hline

        TSO & 
            L $\rightarrow$ L, L $\rightarrow$ S, \newline
            S $\rightarrow$ S &
            S $\rightarrow$ F, F $\rightarrow$ L,
            F $\rightarrow$ F \\ \hline

        PSO & 
            L $\rightarrow$ L, L $\rightarrow$ S &
            S $\rightarrow$ F, F $\rightarrow$ S, F $\rightarrow$ L, 
            \newline
            F $\rightarrow$ F \\ \hline


        RC & 
            & L/S $\rightarrow$ rel,
            acq $\rightarrow$ L/S, \newline
            rel/acq $\rightarrow$ rel/acq
            \\ \hline

        \end{tabular}
    }
    \end{center}
    \label{tab:consistency}
    \vspace{-.2in}
\end{table}

\subsubsection{Partial Store Order (PSO)} \label{sec:tardis-pso}

The PSO consistency model~\cite{weaver1994} relaxes both the Store 
$\rightarrow$ Load and the Store $\rightarrow$ Store orderings.  
Similar to TSO, we use the \lts and \sts to model the program order 
constraints. In PSO, since Load $\rightarrow$ Load is enforced but 
Store $\rightarrow$ Load is not, which is the same as TSO, the rule 
for \lts is also the same as TSO. Namely, \pts should monotonically
increase independently of store timestamps. 

The timestamp order for stores, however, does not need to 
monotonically increase, since Store $\rightarrow$ Store is relaxed.  
Therefore, the timestamp of a store (\ts) only needs to be no less 
than the \lts ($\ts \geq \lts$). And \sts represents the largest 
store timestamp so far (instead of the last store timestamp), namely 
\textit{\sts $=$ Max(\sts, \ts)}. 

For a fence instruction, \lts synchronizes with \sts, namely \lts $=$ 
\textit{Max(\lts, \sts)}. The resulting \lts is the timestamp of the 
fence.

\subsubsection{Release Consistency (RC)} \label{sec:tardis-rc}

Release consistency~\cite{gharachorloo1990} relaxes all the program 
order constraints; furthermore, it also relaxes the ordering 
constraints for synchronizations. Specifically, an \textit{acquire} 
guarantees that all the following (but not the previous) operations 
are ordered after the acquire and a \textit{release} guarantees that 
all the previous (but not the following) operations are ordered before 
the release. 

In Tardis, we need to maintain timestamps for acquire (\acquirets) and  
release (\releasets) operations, as well as the maximal commit 
timestamp (\maxts) so far.  A normal load or store operation 
(commit timestamp \ts) can be performed as long as its timestamp 
is greater than \acquirets (\ts $\geq \acquirets$); \maxts represents 
the largest commit timestamp as seen by the core so far (\maxts $=$ 
\textit{Max}(\maxts, \ts)). At a release instruction, \releasets and 
\maxts are synchronized (\releasets $=$ \textit{Max}(\releasets, 
\maxts)). At an acquire instruction, \acquirets and \releasets are 
synchronized (\acquirets $=$ \textit{Max}(\acquirets, \releasets)).


\section{Proof of Correctness} \label{sec:proof}

In this section, we prove that the algorithm in \cref{sec:tso-impl} 
correctly implements TSO. We first introduce several definitions and 
invariants (in the form of lemmas) of the Tardis protocol 
(\cref{sec:proof-lemma}). We then use these lemmas to prove that our 
algorithm implements TSO (\cref{sec:proof-tso}).

\subsection{Invariants of Tardis protocol} \label{sec:proof-lemma}

The invariants of Tardis introduced in this section are true 
regardless of the memory model used by the system. While some of the lemmas 
come from a proof of Tardis SC (\cite{tardis-proof}), we restate 
them in the language of physiological time. 

To simplify the discussion, we first introduce the concept of 
\textit{master} and \textit{snapshot} cachelines.

\begin{definition}[Master, Snapshot Cacheline] \label{def:master}
A cacheline is a master cacheline if it is in \M state in an L1, or in 
\Sh state in the LLC. A cacheline is a snapshot cacheline if it is in 
\Sh state in an L1.
\end{definition}

Here are some facts about master and snapshot cachelines:

\textbf{Fact 1}. \textit{For each cacheline with (\wts, \rts), the 
data value comes from a previous store that committed at logical time 
\wts.} 

In Tardis, the only way to change \wts of a cacheline is by performing 
a store to it and the new \wts after the store always equals the
commit timestamp of the store.

\textbf{Fact 2}. \textit{For each address in the system, at most one 
master cacheline exists but multiple snapshot cachelines may exist.} 

This is similar to the single-writer, multiple-reader (SWMR) invariant 
in traditional coherence protocols. In Tardis, however, multiple 
snapshot cachelines (i.e., shared L1 cachelines) can coexist with a 
master cacheline (e.g., exclusive L1 cacheline) at the same physical 
time. 

\textbf{Fact 3}. \textit{A snapshot cacheline is derived by taking a 
snapshot of data and timestamps of a master cacheline}. 

In the protocol, an L1 shared cacheline may be derived from a shared 
response from the LLC or from downgrading a modified cacheline in an 
L1. In both cases, it is a snapshot of a master cacheline.  

For the proof, we use (\textit{ts}, \textit{pt}) to represent the 
physiological time of an operation, where \textit{ts} and \textit{pt} 
are the logical and physical commit time of the operation,
respectively.

\begin{lemma} \label{lemma:increase}
For each address, the \wts and \rts of the master cachelines never 
decrease.
\begin{proof} 
In the basic Tardis protocol, no operation decreases the timestamp of 
a cacheline. For a master cacheline, its \wts/\rts can only increase 
(through writes) or stay the same (through exclusive response to L1 or 
write back to LLC) but never decrease.
\end{proof}
\end{lemma}

\begin{lemma} \label{lemma:master}
For a master cacheline, no store to the address has happened at 
$(wts', wpt')$ such that $(wts, wpt)$ $<$ $(wts', wpt')$, where the 
($wts$, $wpt$) is the physiological commit time of the store that 
created the data in the cacheline.
\begin{proof} 
If such a store has ever happened, it would have created a version of 
master cacheline with write timestamp $\wts' > \wts$. However, 
Lemma~\ref{lemma:increase} states that the \wts of the master 
cachelines do not decrease. This contradicts
the fact that the \wts of the cacheline is currently less than 
$\wts'$.
\end{proof}
\end{lemma}

\begin{lemma} \label{lemma:snapshot}
For a snapshot cacheline (with \wts and \rts) at physical time \pt, no 
store has happened at $(\ts, \pt')$ such that $(\wts, \wpt) < (\ts, 
\pt') < (\rts, \pt)$, where the ($\wts$, $\wpt$) is the physiological 
commit time of the store that created the data in the cacheline.
\begin{proof} 
A snapshot cacheline is always copied from a master cacheline. When 
the snapshot is taken, according to Lemma~\ref{lemma:master}, no store 
to the address exists at a physiological time after $(\wts, \wpt)$.  
Since then, according to Tardis rules, a store can only happen at a 
logical timestamp greater than \rts of the cacheline. Therefore, any 
later store has a greater physiological time than $(\rts, \infty)$.  So 
no store can exist between $(\wts, \wpt)$ and $(\rts, \pt)$.
\end{proof}
\end{lemma}

\subsection{Tardis TSO Proof} \label{sec:proof-tso}

We first make the following assumption about a processor implementing 
TSO.

\begin{assumption} \label{assumption:tso}
A TSO processor commits instructions in physical time order if their 
global memory order has to follow the program order according to Rule
1 and 3 of the TSO definition. Specifically, 

\hspace{.2in}L(a) $<_p$ L(b) $\Rightarrow$ L(a) $<_{pt}$ L(b)

\hspace{.2in}L(a) $<_p$ S(b) $\Rightarrow$ L(a) $<_{pt}$ S(b)

\hspace{.2in}S(a) $<_p$ S(b) $\Rightarrow$ S(a) $<_{pt}$ S(b)

\hspace{.2in}X $<_p$ FENCE $\Rightarrow$ X $<_{pt}$ FENCE

\hspace{.2in}FENCE $<_p$ X $\Rightarrow$ FENCE $<_{pt}$ X

\end{assumption}

This assumption does not require extra hardware changes to existing 
processors. In fact, processors implementing TSO today already follow 
this assumption. 

\begin{theorem} \label{thm:main}
The protocol of \cref{sec:tso-impl} implements TSO.

\begin{proof} 
We will prove that each rule in the definition of TSO is maintained in 
our protocol.

\textbf{Proof for Rule 1.} According to \cref{sec:tso-pts}, both
\lts and \sts monotonically increase and for a store, its \sts is no 
less than the current \lts. In other words, 

\hspace{.2in}L(a) $<_p$ L(b) $\Rightarrow$ L(a) $\leq_{ts}$ L(b)

\hspace{.2in}L(a) $<_p$ S(b) $\Rightarrow$ L(a) $\leq_{ts}$ S(b)

\hspace{.2in}S(a) $<_p$ S(b) $\Rightarrow$ S(a) $\leq_{ts}$ S(b)

Combined with our assumption of a TSO processor 
(Assumption~\ref{assumption:tso}), we have:

\hspace{.37in}X $\leq_{ts}$ Y and X $<_{pt}$ Y

\hspace{.2in}$\Rightarrow$ X $<_{ts}$ Y or (X $=_{ts}$ Y and X 
$<_{pt}$ Y) 

\hspace{.2in}$\Rightarrow$ X $<_{pl}$ Y 

\hspace{.2in}$\Rightarrow$ X $<_m$ Y

which finishes the proof of Rule 1 of TSO.

\textbf{Proof for Rule 2.}
We will prove that for each memory load, the returned value is the one 
specified by Rule 2 of the TSO definition. According to Fact 1, for 
each cacheline in Tardis, the data value was created by a store
that happened at the cacheline's \wts. A load to a cacheline observes 
the value of this particular store. Assuming that the store committed 
at physical time \wpt, we need to prove that the physiological 
time of this store, which is (\wts, \wpt), is greater than all the 
other stores in the set $S = S_1 \cup S_2$ where $S_1 = \{S(a) | S(a) 
<_m L(a)\}$ and $S_2 = \{S(a) | S(a) <_p L(a)\}$ and that the observed 
store is also in this set.  Specifically, we consider the following 
two cases.

\underline{Case 1, load to a shared L1 cacheline}. Timestamps of 
the cacheline are \wts and \rts.  Because \wts $\leq$ \lts $\leq$ \rts 
and \wpt $<$ \pt, the observed store must be in the set $S_1$.  
According to Lemma~\ref{lemma:snapshot}, at physical time \pt, no 
store has happened at ($\wts'$, $\wpt'$)  such that (\wts, \wpt) < 
(\wts$'$, \wpt$'$) < (\rts, \pt). As a result, the observed store has 
the largest physiological time in set $S_1$. 

We now prove that all stores in set $S_2$ have smaller physiological 
time than the observed store. This can be proven by contradiction.  If 
such a store has been executed by the current core at ($\wts'$, 
$\wpt'$) > (\wts, \wpt), the current core must have owned a master 
cacheline with $\wts'$ after the store. Since the master cacheline's 
\wts never decreases and a snapshot cacheline is copied from a master
cacheline, the current shared cacheline must have \wts greater than 
$\wts'$, contradicting the assumption. So the observed store has the 
largest physiological time in set $S_1 \cup S_2$, proving the 
invariant. 

\underline{Case 2, load to a modified L1 cacheline}. In this case, the 
loaded cacheline is a master cacheline. If the cacheline is not dirty, 
then \wts $\leq$ \lts.  So the observed store is in $S_1$.  If the 
cacheline is dirty, then the observed store must be before the current 
load in the program order and is therefore in $S_2$. According to 
Lemma~\ref{lemma:master}, no store has happened to the address at 
($\wts'$, $\wpt'$) > (\wts, \wpt). So the observed store must have the 
largest physiological time in set $S = S_1 \cup S_2$.
 
\textbf{Proof for Invariant 3.} In Tardis TSO, a fence synchronizes 
\lts and \sts. Tardis TSO enforces the following:

X $<_p$ FENCE $\Rightarrow$ X $\leq_{ts}$ FENCE

X $<_p$ FENCE $\Rightarrow$ X $<_{pt}$ FENCE 

FENCE $<_p$ X $\Rightarrow$ FENCE $\leq_{ts}$ X

FENCE $<_p$ X $\Rightarrow$ FENCE $<_{pt}$ X

Combining these equations and applying the definition of physiological 
time, we prove the last invariant of TSO.
\end{proof}

\end{theorem}

\section{Renewal Reduction Techniques} \label{sec:renew}

As discussed in \cref{sec:background}, a major drawback of the 
original Tardis protocol is the renewal problem. With load 
speculation, latency of renew messages can be largely hidden, but 
network traffic overhead remains. A renew message is unnecessary if 
the renewed line is unchanged. In this section, we discuss techniques 
to reduce unnecessary renew messages.  
 
\subsection{MESI on Tardis}

The original Tardis protocol implements MSI where a read to a private 
cacheline loads it in S (shared) state in the L1 cache. As the \lts 
(or \pts in SC) increases in a core, these data will expire and be 
renewed. These renewals are unnecessary since there is no need to 
maintain coherence for a core's private data.  

The MESI protocol can resolve this issue. MESI adds an E (Exclusive) 
state to MSI. The E state is granted for a request if the cacheline is 
not shared by other cores. Like a cacheline in M state, an exclusive 
cacheline in a private cache never expires. If the \lts is greater 
than the \rts, the \rts of that line can be increased to \lts with no 
renewal. This can be done since the line is not shared by other cores, 
namely, a \textit{master cacheline} according to 
Definition~\ref{def:master}. Therefore, renewal does not happen for 
private data in MESI.  

Different from traditional coherence protocols, Tardis can grant E 
state to a core even if other cores are still sharing the line. This 
is similar to granting M state without the need of invalidation.  
However, for performance reasons, it is still desirable to only grant 
E state for private data. In Tardis, a cacheline is likely to be not 
shared if it has just been loaded from DRAM, and if it has just been 
downgraded from E/M to S state.  Therefore, we add an E-bit to each 
cacheline in the LLC to indicate whether the cacheline is likely to be 
shared or not. The E-bit is set when the cacheline is loaded from DRAM 
and also when the cacheline is downgraded to S state.  The E-bit is 
reset when the cacheline becomes cached upon load request.  Note that 
E-bit may be unset even if no core is sharing the line (\eg, all 
sharers silently evict the line).  This does not affect the 
functional correctness of the implementation. 

\subsection{Livelock Detection} \label{sec:livelock}

A disadvantage of Tardis is that propagating a store to other cores 
may take an arbitrarily long time. This is because the writer does not 
notify the 
sharing cores who may not \textit{pull} the latest value if they keep
reading the stale version.  In the worst case if a core spins on a 
stale cacheline with a small \lts (or \pts in SC), it never sees the 
latest update and livelock occurs (\cref{lst:spin}). Although such 
livelock is not precluded by the consistency model, it should be 
disallowed by the coherence protocol which requires that every write 
should eventually propagate to other cores. In practice, this spinning 
behavior is commonly used for communication in parallel programs.

\vspace{-.05in}
\begin{lstlisting}[language=C,label={lst:spin},caption={Threads 
communicate through spinning}, escapeinside={@}{@}]
    [Core 0]            [Core 1]@\vspace{0.05in}@
   @//\textit{spin (lts = 10)}@       @//\textit{store (sts = 20)}@
   while (!done) {      ...
     sleep(1)           done = true
   }                    ...
\end{lstlisting}
\vspace{-.05in}

\subsubsection{Baseline: Periodic Self Increment}

The original Tardis protocol solves the livelock problem by self 
incrementing the \pts (or \lts in TSO) periodically to force the 
logical time in each core to move forward. For a spinning core (e.g., 
core 0 in \cref{lst:spin}), the \lts will increase and eventually 
become greater than the \rts of the cacheline being spun on at which 
point the line expires and the latest value will be loaded.  

Frequently self incrementing \lts causes two performance issues.  
First, the shared cachelines in the private cache may frequently 
expire,  generating renewals. For programs without spinning, these 
renewals are unnecessary but incur network traffic and latency.  
Second, for cachelines that have an \rts much larger than the current 
\lts of
the core, it may take a significant amount of time before the \lts 
increases to \rts for the stale line to expire.


\subsubsection{Livelock Detector}

We make a key observation that a \textit{check} message can be sent to 
check the freshness of a cacheline even before the cacheline actually 
expires. 
Like a renew request, if the latest data is newer than the cached 
data, the latest cacheline is returned.  If the cached data is already 
the latest version, however, a check response is returned without 
extending the \rts of the cacheline in the LLC. 

The check request can resolve both drawbacks of the self increment
scheme. Since a core does not need to frequently increase its \lts, 
the number of renewals can be reduced. Also, a check request can be 
sent when \lts is much smaller than \rts, so a core does not need to 
wait a long time for the cacheline to expire. 
 
Generally, a check request should be sent when the program seems to 
livelock as it keeps loading stale cachelines. In practical programs, 
such a livelock is usually associated with variable spinning, which 
typically involves a small number of cachelines and is therefore easy 
to detect in hardware.


\begin{algorithm}
  \caption{Livelock Detection Algorithm (called for each read request 
  to a shared L1 cacheline).} \label{alg:livelock}
  \begin{algorithmic}[1]
    \footnotesize
    \STATE \textbf{Input:} addr \ \ \ \ \  \textit{// memory access 
    address}
    \STATE \textbf{Return Value:} whether to issue check request
    \STATE \textbf{Internal State:} AHB, thresh\_count 
    
    \vspace{.1in}
    \IF{AHB.contains(addr)}
        \STATE AHB[addr].access\_count ++
        \IF{AHB[addr].access\_count == thresh\_count}
            \STATE AHB[addr].access\_count = 0
            \STATE \textbf{return} true
        \ENDIF
    \ELSE
        \STATE AHB.enqueue(addr) 
        \STATE AHB[addr].access\_count = 0
        \STATE \textbf{return} false
    \ENDIF
 \end{algorithmic}
\end{algorithm}
 
We designed a small piece of hardware next to each core to detect 
livelock. 
It contains an \textit{Address History Buffer} (AHB) and a threshold 
counter (\textit{thresh\_count}). The AHB is a circular buffer keeping 
track of the most recently loaded addresses.  Each entry in AHB 
contains the address of a memory access, and an \textit{access\_count},
which is the number of accesses to the address since it was loaded to 
AHB. When \textit{access\_count} becomes greater than the 
\textit{thresh\_count}, a check request is sent for this address 
(Algorithm~\ref{alg:livelock}).  The value of
\textit{thresh\_count} can be static or dynamic.  We chose to use an 
adaptive threshold counter scheme 
(Algorithm~\ref{alg:threshold-counter}) in order to minimize the 
number of unnecessary check messages.  

\begin{algorithm}
  \caption{Adaptive Threshold Counter Algorithm (called for each check 
  response).  } \label{alg:threshold-counter}
  \begin{algorithmic}[1]
    \footnotesize
    \STATE \textbf{Input:} check\_update \textit{// whether the 
    checked address has been updated.}
    \STATE \textbf{Internal State:} thresh\_count
    \STATE \textbf{Constant:} min\_count, max\_count, check\_count, 
    check\_thresh \vspace{.1in}
    
    \IF{check\_update}
        \STATE thresh\_count = min\_count
        \STATE check\_count = 0
    \ELSE
        \STATE check\_count ++
        \IF{check\_count == check\_thresh \textbf{and}\\ thresh\_count 
        < max\_count}
            \STATE thresh\_count = thresh\_count $\times$ 2
        \ENDIF
    \ENDIF
  \end{algorithmic}
\end{algorithm}

The livelock detection algorithm (Algorithm~\ref{alg:livelock}) is 
executed when reading a shared cacheline.  It is not called when  
accessing cachelines in E or M state since no livelock can occur for 
those accesses.  If the accessed address does not exist in the AHB, a 
new entry is allocated. Since AHB is a circular buffer, this may evict 
an old entry from it. We use the LRU replacement policy here but other 
replacement policies should work equally well. For an AHB hit, the 
\textit{access\_count} is incremented by 1.  If the counter saturates 
(i.e., reaches \textit{thresh\_count}), a check request is sent and 
the \textit{access\_count} is reset.  All \textit{access\_counts} are 
reset to 0 when the \lts increases due to a memory access, since this 
indicates that the core is not livelocking, and thus there is no need 
to send checks.  

The counter \textit{thresh\_count} may be updated for each check 
response (Algorithm~\ref{alg:threshold-counter}). If the checked 
address was updated by another core, then \textit{thresh\_count} 
should decrease to the minimal value, indicating that check requests 
should be sent more frequently since data seems to be updated 
frequently. Otherwise, if \textit{check\_thresh} number of consecutive 
check requests
returned without data being changed, then \textit{thresh\_count} is 
doubled since it appears unnecessary to send check requests that 
often.  Adaptively determining the value of \textit{thresh\_count} can 
reduce the number of unnecessary check requests if a thread needs to 
spin for a long time before the data is updated.

Note that the livelock detector can only detect spinning involving 
loads to less than $M$ (the number of entries in AHB) distinct 
addresses. So in theory, the livelock detector cannot capture all 
possible livelocks and self incrementing \lts is still required to 
guarantee forward progress.  For practical programs, however, spinning 
typically involves a small number of distinct addresses. So the 
livelock detector is able to capture livelock in the vast majority of 
programs.  We still self increment \lts periodically but the frequency 
can be much lower, since most programs' performance does not rely on 
this mechanism anymore.

\subsection{Lease Prediction}

During regular operation of Tardis, memory stores are the main reason 
that the timestamps increase in the system. The amount that an \sts 
increases is determined by the lease of the previous data version, 
because the \sts of the store must be no less than the cacheline's 
previous \rts. Therefore, the lease of each cacheline is important to 
the timestamp incrementing rate as well as the renew rate. The 
original Tardis protocol uses a static lease for every shared 
cacheline. We first show that a static leasing policy may incur 
unnecessary renewals (\cref{sec:static-lease}). We then propose a 
dynamic leasing policy to mitigate the problem 
(\cref{sec:lease-predictor}).

\subsubsection{Static Lease vs. Dynamic Lease} 
\label{sec:static-lease}

In the code snippet shown in Listing~\ref{lst:lease}, both cores run 
the same program. They both load addresses A and B and then store to 
B. When the cachelines are loaded to L1 caches, they are all reserved 
with a static lease $L$. When the store to address B is performed, 
both cores' \sts  
jump ahead by at least $L$. 
At the end of the loop, the FENCE instruction increases \lts to the 
value of \sts.

\vspace{-.1in}
\begin{lstlisting}[language=C,label={lst:lease},caption={The case 
study parallel program}, escapeinside={@}{@}]
  [Core 0]              [Core 1] @\vspace{0.05in}@
   while(B < 10) {      while(B < 10) {
     print A               print A
     B++                   B++
     FENCE                 FENCE
   }                     }
\end{lstlisting}

In the next iteration when both cores load A again, the cacheline has 
expired in the L1 caches. Because the \lts has already jumped ahead by 
$L$ due to the previous store B and the fence, but the lease on A was 
also $L$.  As a result, in each iteration of the loop, \lts and \sts 
jump ahead by $L$ and A is renewed at each core.  All these renewals 
to A are successful, and therefore unnecessary, since A has never been 
changed. Note that using a larger static $L$ does not solve the 
problem since \lts/\sts will jump ahead further and A will still 
expire. 

Our solution to this problem is to use different leases for different 
addresses. Intuitively, we want to use large leases for read only or 
read intensive data, and use small leases for write intensive data.  
In the example in Listing~\ref{lst:lease}, if A has lease 100 and B 
has lease 10, then each store to B increases the \sts and \lts only by 
10. So it takes about 10 iterations before A has to be renewed again.  
The renew rate is mainly a function of the ratio between these two 
leases; the absolute lease values are not critical.

In a real system, it is non-trivial to decide what data should have a 
larger or smaller lease. Here, we explore hardware only solutions and 
design a predictor to decide the lease for each cacheline. It is 
possible to do this more accurately with software support, such 
explorations are left for future work.  

\subsubsection{Lease Predictor}
\label{sec:lease-predictor}

Our lease predictor is based on the observation that cachelines that 
are frequently renewed are more likely to be read intensive.  
Therefore, a cacheline should be reserved with a larger lease if it is 
renewed frequently. The logic to determine the lease value is built 
into the LLC which has knowledge of all the renewals. 

For each renew request from the L1 to the LLC, the last lease 
(\textit{req\_lease}) of the cacheline is also sent to the LLC and is 
processed by the lease predictor. For an L1 miss, the cacheline has no 
last lease so the \textit{min\_lease} is used. The lease predictor 
computes a proper lease based on the \textit{req\_lease} and the 
predictor's current internal lease (\textit{cur\_lease}).  
Specifically, the algorithm of our lease predictor is shown in 
Algorithm~\ref{alg:lease}.

\begin{algorithm}
  \caption{Lease Prediction Algorithm (called for each LLC request).} 
  \label{alg:lease}
  \begin{algorithmic}[1]
    \footnotesize
    \STATE \textbf{Input}
    \STATE \hspace{0.2in}req\_lease \ \ \  \textit{// lease in the 
    request}
    \STATE \hspace{0.2in}req\_type \ \ \ \ \textit{// WRITE, READ or 
    RENEW}
    \STATE \textbf{Return Value:} Lease of the returned cacheline.
    \STATE \textbf{Internal State:} cur\_lease
    \vspace{.1in}
    \IF{req\_type == WRITE}
        \STATE cur\_lease = min\_lease
    \ELSIF{req\_type == RENEW \textbf{and} req\_lease == cur\_lease \\ 
            \hspace{.15in}\textbf{and} cur\_lease $<$ max\_lease}
            \STATE cur\_lease = cur\_lease $\times$ 2
    \ENDIF
    \STATE \textbf{return} cur\_lease
 \end{algorithmic}
\end{algorithm}

For a write request, the \textit{cur\_lease} is updated to the minimal 
lease value (\textit{min\_lease}). Our reasoning is that the write indicates that 
the cacheline might be write intensive and so assigning a large lease 
to it may cause unnecessary renewals of other cachelines. For a normal 
read request, \textit{cur\_lease} is used for the requested 
cacheline. For a renew request, \textit{cur\_lease} is compared 
with the request lease (\textit{req\_lease}). If they are different, 
then \textit{cur\_lease} is used for the cacheline.  Otherwise, 
\textit{cur\_lease} is doubled since the cacheline seems to be 
renewed multiple times by the same core and is therefore likely to be 
read intensive. If \textit{cur\_lease} already reached the maximal 
value (\textit{max\_lease}), then it should no longer increase.

The initial value of \textit{cur\_lease} is the minimal value 
(\textit{min\_lease}). This means that for a cacheline first loaded to 
the LLC, we always assume it to be write intensive. We made this 
design decision because incorrectly giving a large lease to a write 
intensive cacheline is much more harmful than giving a small lease
to a read intensive cacheline. If a cacheline with a large lease is 
written, a large number of cachelines in the core's L1 might expire 
due to the program timestamp jumping ahead. In contrast, if a read 
only cacheline is given a small lease, only this cacheline needs to be 
renewed and other cachelines are not affected. 

\section{Evaluations} \label{sec:eval}

We evaluate the performance of Tardis with relaxed 
consistency models and the optimizations proposed in \cref{sec:renew}. 

\subsection{Methodology}

\vspace{-.05in}
\subsubsection{System Configuration}

We use the Graphite~\cite{graphite} multicore simulator to model the
Tardis coherence protocol. Configurations of the hardware, Tardis, and
its enhancements are shown in \cref{tab:system} and \cref{tab:tardis}.

For the baseline Tardis, we implemented all the optimizations in the 
original Tardis protocol, including speculative reads when a cacheline 
expires in the L1 cache and not incrementing \sts for private writes 
\cite{tardis}.  The static lease always equals 8. And the \lts self 
increments by 1 for every 100 memory accesses.

For the livelock detector, the address history buffer (AHB)  contains 
8 entries by default. The threshold counter can take values ranging 
from 100 to 800. The threshold counter is doubled if 10 consecutive 
checks respond that the data has not been changed.

The minimal lease is chosen to be 8 and the maximum lease is 64. There 
are four possible lease values (i.e., 8, 16, 32, 64) in the system.  
We chose four lease values because having more values does not 
significantly affect performance.   

\begin{table}
    \caption{ System Configuration.}
	\vspace{-.05in}
	\begin{center}
    { \footnotesize
        \begin{tabular}{|l|l|}
            \hline
			\multicolumn{2}{|c|}{System Configuration} \\
            \hline
            Number of Cores				& N = 64 \\
            Core Model                  & Out-of-order, 128-entry ROB 
            \\ 
			\hline
            \hline
			\multicolumn{2}{|c|}{Memory Subsystem} \\			
            \hline
			Cacheline Size        		& 64~bytes \\
			L1 I Cache					& 16~KB, 4-way \\
			L1 D Cache                	& 32~KB, 4-way \\
			Shared L2 Cache per Core    & 256~KB, 8-way \\
			DRAM Bandwidth				& 8 MCs, 10~GB/s per MC\\
			DRAM Latency				& 100~ns\\
			\hline
			\hline
			\multicolumn{2}{|c|}{2-D Mesh with XY Routing} \\
            \hline Hop Latency					& 2 cycles (1-router,
										1-link)\\
			Flit Width					& 128 bits\\
			\hline
        \end{tabular}
    }
	\end{center}
    \label{tab:system}
\end{table}

\begin{table}
    \vspace{-.15in}
    \caption{ Tardis Configuration. }
	\vspace{-.05in}
	\begin{center}
    { \footnotesize
        \begin{tabular}{|l|l|}
            \hline
            \multicolumn{2}{|c|}{Baseline Tardis} \\
            \hline
            Static Lease 				& 8 \\
            Self Increment Period       & 100 memory accesses\\
            \hline
            \hline
            \multicolumn{2}{|c|}{Livelock Detector} \\
            \hline
            AHB size            		& 8 entries \\
            Threshold Counter   		& min 100, max 800\\
            Check Threshold             & 10 \\
            \hline
			\hline
            \multicolumn{2}{|c|}{Lease Prediction} \\
			\hline
            Minimal Lease Value         & 8 \\
            Maximal Lease Value         & 64 \\
            \hline
        \end{tabular}
    }
	\end{center}
    \label{tab:tardis}
\end{table}

\vspace{-.05in}
\subsubsection{Baselines}

The following coherence protocols are implemented and evaluated for 
comparison. Except in \cref{sec:eval-consistency}, all the 
configurations use the TSO consistency model and MESI.  

\textbf{Directory:} Full-map MESI directory coherence protocol.

\textbf{Base Tardis:} Baseline Tardis~\cite{tardis} where \lts self 
increments by 1 for every 100 memory accesses. 

\textbf{Tardis + live:} Baseline Tardis with livelock detection. \lts 
self increments by 1 for every 1000 memory accesses.

\textbf{Tardis + live + lease:} Tardis with both a livelock detector and 
lease predictor.  

In Tardis, the base-delta timestamp compression
scheme~\cite{tardis} was implemented and each timestamp requires 20 
bits of storage. This means that each cacheline requires 40 bits in total for \wts and 
\rts, regardless of the number of cores in the system. The full-map 
directory protocol, in contrast, requires an $N$-bit sharer list for 
each cacheline in the LLC for an $N$-core system.
However, note that timestamps are required in the L1 for the baseline and optimized Tardis
unlike in the directory protocol.

\begin{figure}[t!]
	\centering
    \subfloat[Speedup] {
    \includegraphics[width=.45\columnwidth]{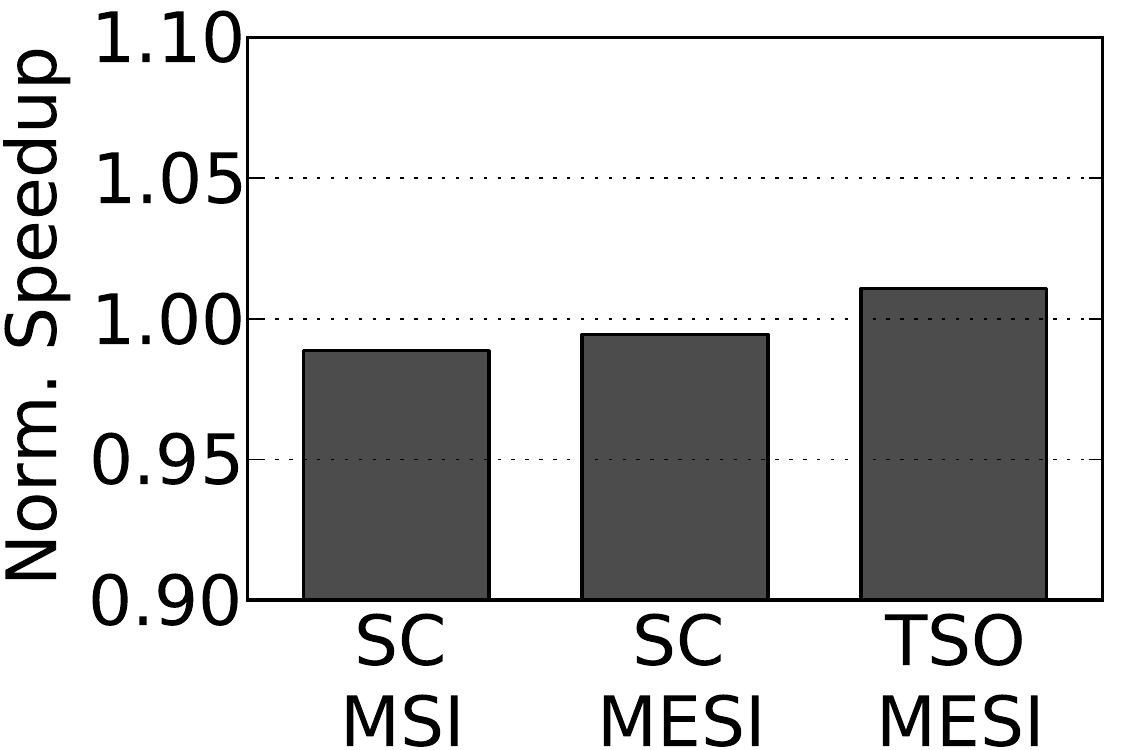}
	\label{fig:eval-tso-thr}
    }
    \subfloat[Renew Rate] {
    \includegraphics[width=.45\columnwidth]{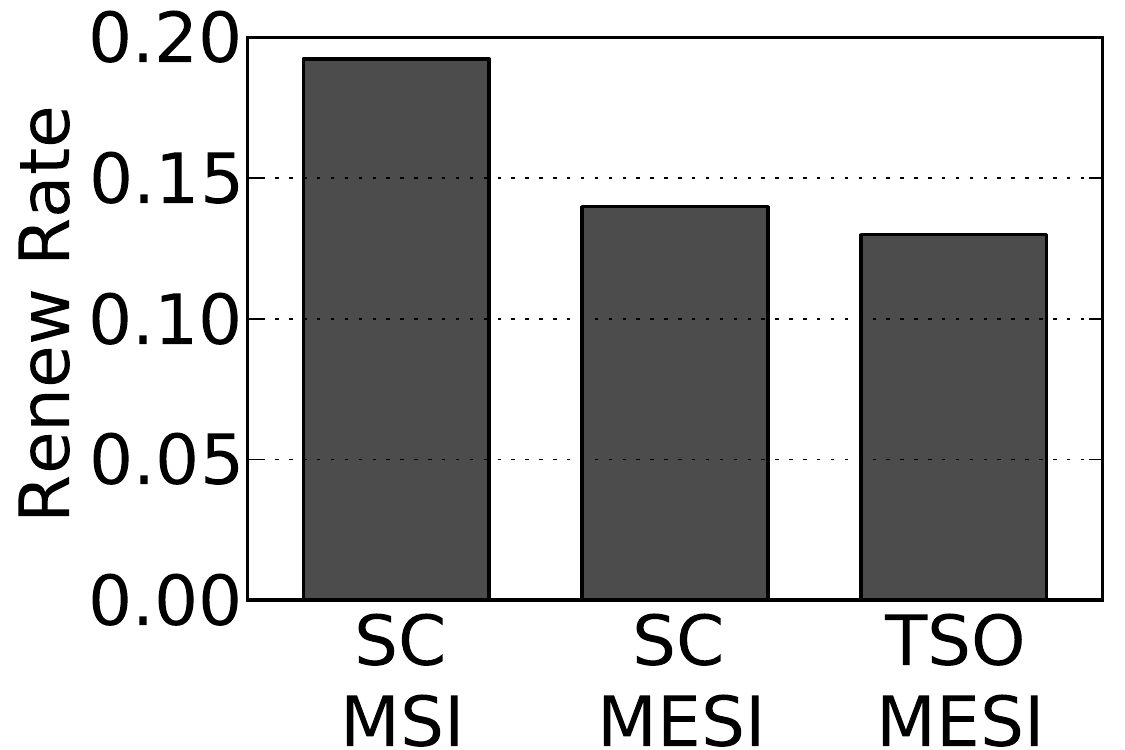}
	\label{fig:eval-tso-renew}
    }
    \caption{ MESI and TSO. Average speedup (normalized to directory + 
    MESI + TSO) and renew rate over all benchmarks. }
	\vspace{-.15in}
\label{fig:eval-tso}
\end{figure}


Our experiments are executed over 22 benchmarks selected from  
Splash2~\cite{splash2}, PARSEC~\cite{parsec}, sparse linear 
algebra~\cite{dongarra2013} and OLTP database 
applications~\cite{yu2014}. For sparse linear algebra, we evaluated 
sparse matrix multiplicaion (\texttt{SPMV}) and symmetric Gauss-Seidel 
smoother (\texttt{SYMGS}) from the HPCG benchmark suite 
(Top 500 supercomputer ranking). For OLTP database, we evaluated two 
benchmarks \texttt{YCSB} and \texttt{TPCC}. All benchmarks are 
executed to completion.    

\subsection{Consistency Models and MESI} \label{sec:eval-consistency}

\cref{fig:eval-tso} shows the speedup of MESI and TSO on Tardis 
normalized to the baseline directory coherence protocol with MESI and 
TSO. All experiments have both livelock detection and lease prediction 
enabled for a fair comparison. Both MESI and TSO can improve the 
overall performance of Tardis. On average, using MESI instead of MSI 
improves the performance of Tardis SC by 0.6\%; using TSO instead of 
SC further improves performance by 1.7\%.

MESI and TSO can also reduce the renew rate of Tardis. We define renew 
rate as the ratio of the number of renew requests over the total 
number of LLC accesses.  \cref{fig:eval-tso-renew} shows the renew 
rate reduction of MESI and TSO on Tardis. MESI reduces the number of 
renew messages since private readonly data is always cached in E state 
and therefore never renewed. TSO further reduces the renew rate since 
the \lts may increase slower than the \pts in SC (cf.  
\cref{sec:tso-pts}) leading to fewer expirations/renewals. MESI and 
TSO together can reduce the average renew rate from 19.2\% to 13.0\%.


Although not shown in these figures, TSO can also significantly 
decrease the rate at which timestamps increase. This is because \lts 
can stay behind \sts. Therefore, \lts and \sts may increase slower 
than how \pts increases in Tardis SC. On average, the timestamp 
increment rate in Tardis TSO is 78\% of the rate in Tardis SC.  

\begin{figure}[t!]
	\centering
	\includegraphics[width=.95\columnwidth]{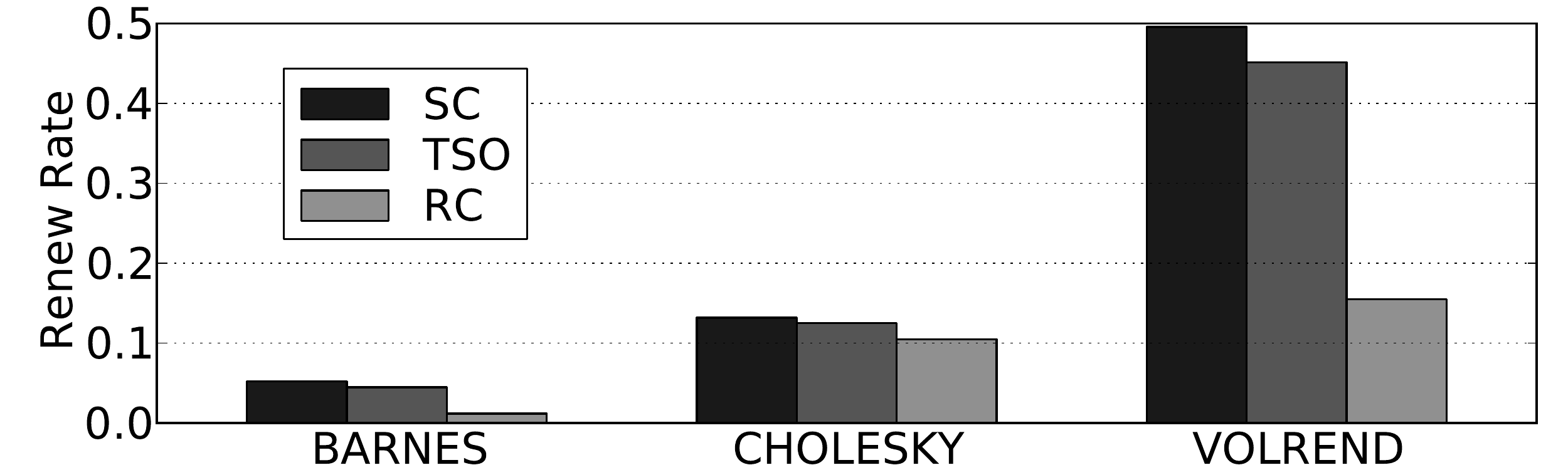}
	\caption{ Renew rate of consistency models SC, TSO and 
RC.  }
\label{fig:consistency}
	\vspace{-.2in}
\end{figure}

\cref{fig:consistency} shows the renew rate of SC, TSO and RC on 
Tardis with MESI. Due to some issues with running x86 pthread 
synchronization primitives on RC, we implemented hardware-based 
synchronization for this experiment and therefore use a separate graph 
to present the results. For many benchmarks that are 
\textit{data-race-free}, relaxing the consistency models does not 
significantly reduce the renew rate. For some other benchmarks (like 
the ones in \cref{fig:consistency}), however, more relaxed consistency 
models lead to significantly fewer renewals.

\subsection{Livelock Detector and Lease Predictor}

\begin{figure*}[t!]
	\centering
    \includegraphics[width=.95\textwidth]{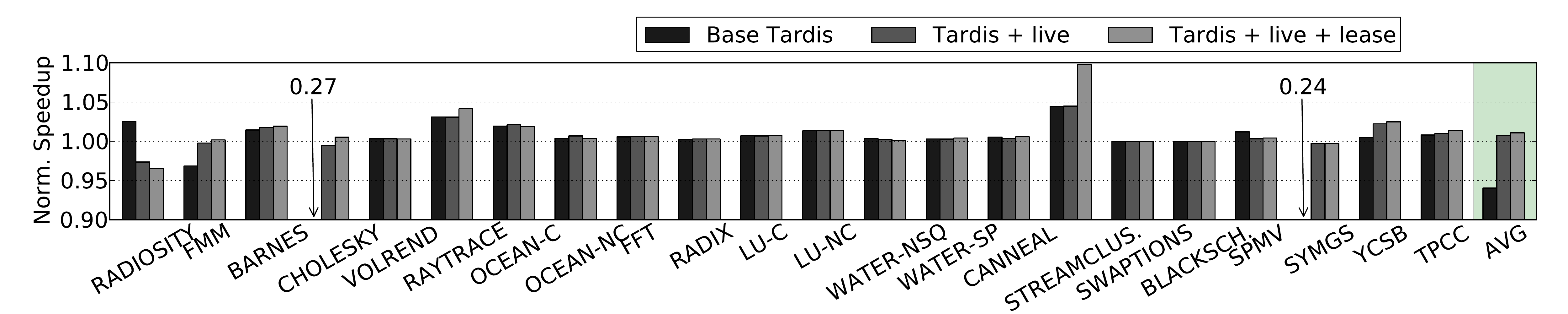}
    \caption{ Speedup of livelock detection and lease prediction 
optimizations. All results normalized to the directory baseline. }
\label{fig:main}
\end{figure*}

\begin{figure*}[t!]
	\centering
    \includegraphics[width=.95\textwidth]{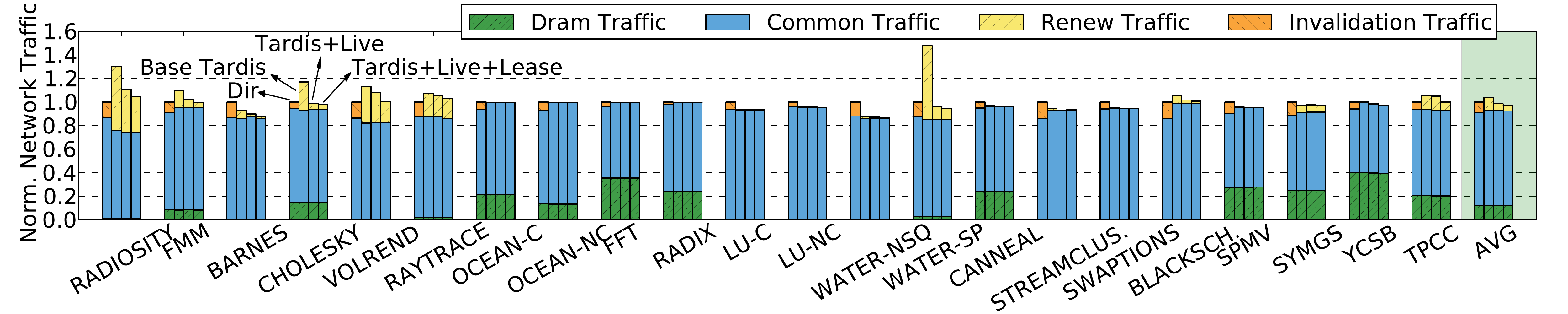}
    \caption{ Network Traffic breakdown of directory and 
different Tardis configurations. All results normalized to the 
directory baseline. }
\label{fig:traffic}
\end{figure*}

We now evaluate the performance and hardware overhead of the livelock 
detector and lease predictor of \cref{sec:renew}. All coherence 
protocols in this section use MESI and TSO.  

\vspace{-.05in}
\subsubsection{Performance and Network Traffic}

\cref{fig:main} shows the performance after adding livelock detection 
and lease prediction compared to the baseline Tardis protocol. All 
numbers are normalized to the baseline directory protocol.

First, we see that \texttt{CHOLESKY} and \texttt{SYMGS} on baseline 
Tardis have much worse performance than the directory protocol.  
Both benchmarks {\em heavily} use spinning to communicate between
threads.  As a result, it may take a long time for the cacheline spun 
on to expire.  The livelock detector can close the performance gap 
between Tardis and directory because a spinning core is able to 
observe the latest data much earlier. \texttt{RADIOSITY} also uses 
spinning. However, the core does other computation between  checking 
the value of the variable. Therefore, our livelock detector cannot 
capture such spinning behavior and forward progress is enforced 
through self incrementing \lts. This leads to suboptimal performance.  
Completely eliminating such performance degradation requires rewriting 
the application using synchronization primitives better than spinning.  
Over the benchmark set, the optimizations improve the performance of 
Tardis by 7.5\% with respect to the baseline Tardis and 1.1\% (up to 
9.8\%) with respect to baseline directory protocol.




\cref{fig:traffic} shows the network traffic breakdown for the same 
four configurations as in \cref{fig:main}. For each experiment, we 
show \textit{dram traffic}, \textit{common traffic}, \textit{renew 
traffic} and \textit{invalidation traffic}. \textit{Common traffic} is 
the traffic in common for both directory coherence and Tardis, 
including shared, exclusive and write back memory requests and 
responses. \textit{Renew traffic} is specific to Tardis including 
renew and check requests and responses. \textit{Invalidation Traffic} 
is specific to the directory-based protocol, including the 
invalidation requests to shared copies from the directory, as well as 
the messages sent between L1 and LLC when a shared cacheline is 
evicted. 

Compared to a directory protocol, Tardis is able to remove all the 
invalidation traffic. However, the renew traffic adds extra overhead.  
The baseline Tardis configuration incurs a large amount of renew 
traffic on some benchmarks (e.g., \texttt{RADIOSITY}, 
\texttt{CHOLESKY}, \texttt{VOLREND} and \texttt{WATER-SP}). Some of 
the renew traffic is due to fast self incrementing \lts (e.g., 
\texttt{RADIOSITY}, \texttt{CHOLESKY} and \texttt{WATER-SP}). For 
these benchmarks, the livelock detection scheme can significantly 
reduce the self increment rate and therefore reduce the amount of 
renew traffic. On average, the livelock detection algorithm
reduces the total network traffic by 5.4\% compared to the baseline 
Tardis.

\begin{figure}[t!]
    \centering
    \includegraphics[width=0.95\columnwidth]{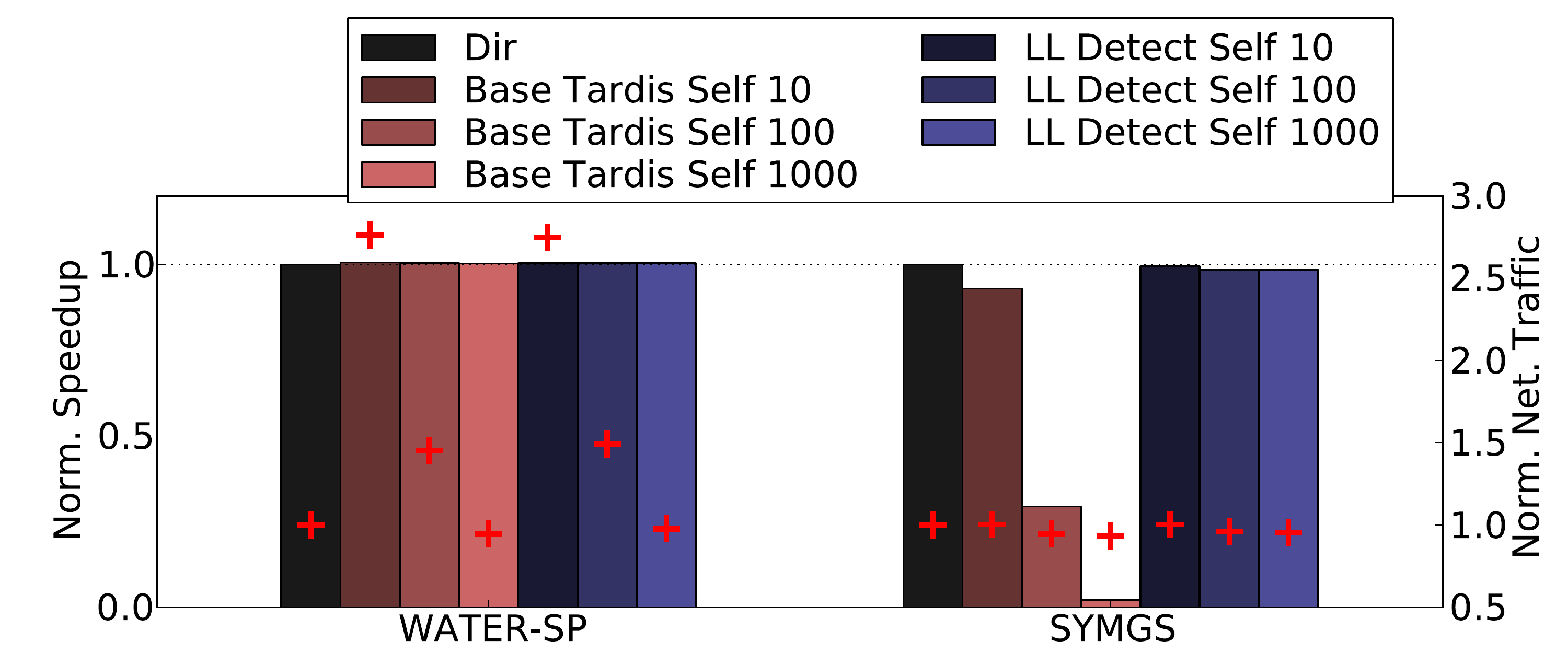}
    \caption{ Sweep the \lts self increment rate.}
    \vspace{-.1in}
\label{fig:selfincr}
\end{figure}


For some benchmarks, shared cachelines expire because the \lts jumps 
ahead due to a write (e.g., \texttt{VOLREND}, \texttt{RADIOSITY}) and 
renew messages are generated. Our lease prediction algorithm is able 
to reduce these unnecessary renewals by using a larger lease for read 
intensive cachelines. On top of the livelock detection optimization, 
lease prediction further reduces the total network traffic by 1.5\% on 
average. With both livelock detection and lease prediction, Tardis can 
reduce the total network traffic by 2.9\% (up to 12.4\%) compared to 
the baseline directory protocol. 

Although not shown here, we also evaluated an idealized leasing scheme 
which is modeled by giving each renew message zero overhead.  The 
idealized scheme has similar performance as the optimized Tardis but 
eliminates almost all the renewal messages; some renewals are still 
needed if the data has actually been changed.

\vspace{-.05in}
\subsubsection{Hardware Complexity}


The hardware overhead for the livelock detector and lease predictor is 
generally small. Each livelock detector contains 8 AHB entries and 
each entry requires an address and a counter. Assuming 48-bit address 
space and a counter size of 2 bytes, the detector only requires 64 
bytes of storage per core. 

To implement the lease predictor, we need to store the current lease 
for each LLC and L1 cacheline. The lease is also
transferred for each shared request and response. However, there is no 
need to store or transfer the whole lease value. Since a lease can 
only take a small number of possible values (e.g., 4 in our 
evaluation), we can use a smaller number of bits (e.g., 2 bits) to 
encode a lease, and the storage overhead is less than 
0.4\% in the LLC and in each L1.   

\begin{figure}[t!]
	\centering
    \includegraphics[width=0.95\columnwidth]{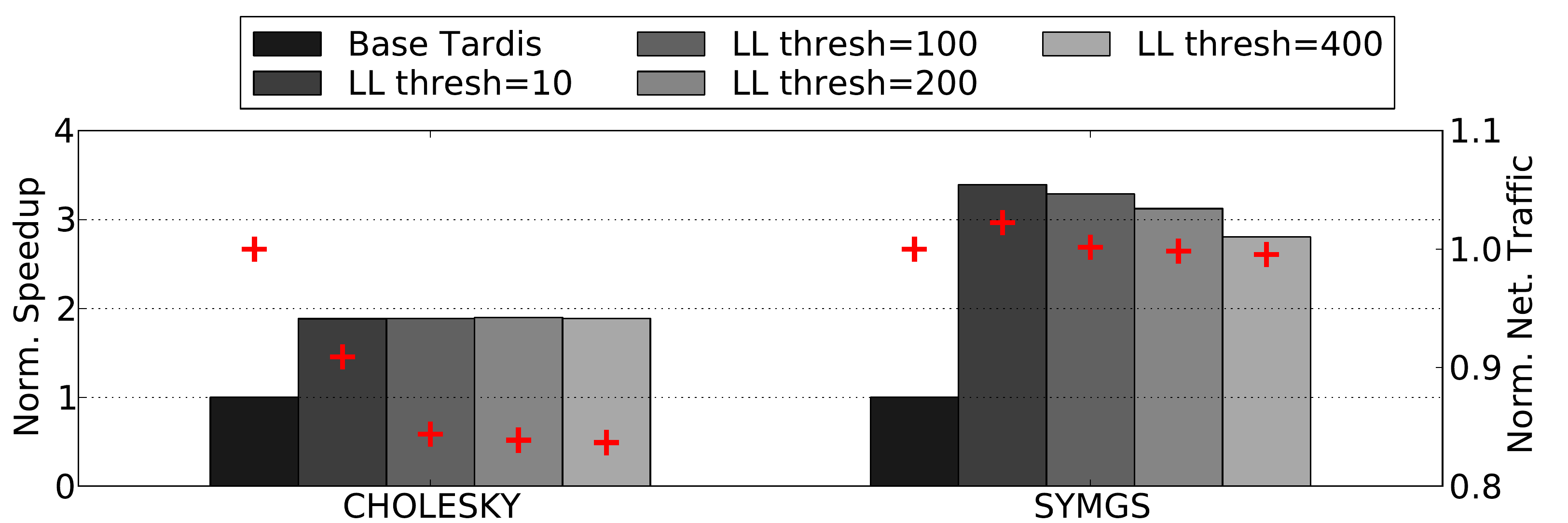}
    \caption{ Sweep the livelock check threshold 
(\textit{thresh\_count}).}
    \vspace{-.1in}
\label{fig:thresh}
\end{figure}

\begin{figure*}[t!]
    \centering
    \includegraphics[width=0.95\textwidth]{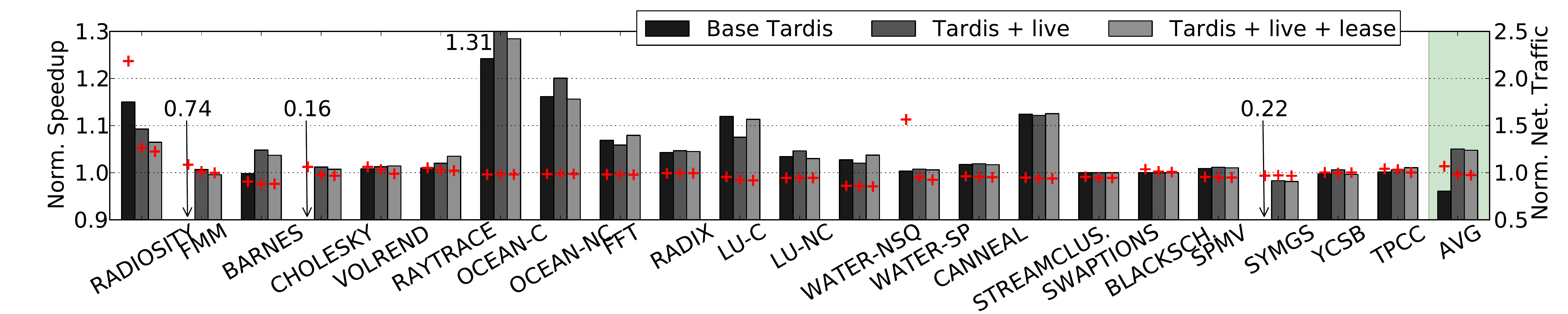}
    \caption{ Tardis vs. Directory on 256 cores.}
\label{fig:256}
\end{figure*}

\subsection{Sensitivity Study}

\vspace{-.05in}
\subsubsection{Self Increment Rate}

\cref{fig:selfincr} shows the performance and network traffic of 
Tardis sweeping the self increment period with and without livelock 
detection. All numbers are normalized to a baseline directory 
protocol. The \texttt{Base Tardis Self 100}  is  the baseline Tardis 
configuration and \texttt{LL Detect Self 1000} is the default 
optimized Tardis configuration (\texttt{LL Detect} stands for livelock 
detect).  

In \texttt{WATER-SP}, changing the self incrementing rate does not 
affect performance regardless of whether livelock detection is turned on 
or not. This is because \texttt{WATER-SP} does not have spinning and 
renewals are unnecessary. Having a large self 
increment period significantly reduces the number of unnecessary 
renewals as well as the total network traffic. 

In \texttt{SYMGS}, for Tardis without livelock detection, performance 
is very sensitive to the self increment period because \texttt{SYMGS} 
intensively uses spinning to communicate between threads. If self 
increment is less frequent, a thread waits longer for the stale data 
to expire and thus performance degrades.  With livelock detection, 
however, check requests are sent when spinning (potential livelock) 
is detected. Therefore, the latest value of a cacheline spun on can be 
returned much earlier. Regardless of the self increment period, Tardis 
with the livelock detector can always match the performance of the 
baseline directory protocol. 

\vspace{-.05in}
\subsubsection{Address History Buffer Size}

We swept the number of entries in the address history 
buffer in a livelock detector for \texttt{CHOLESKY}  and 
\texttt{SYMGS}. According to the results (not shown), as long as the AHB buffer 
size is no less than 2, performance does not change. This is because 
in both (and most other) programs, spinning only involves a very small 
number of distinct memory addresses. \texttt{CHOLESKY} only spins on 
two addresses and \texttt{SYMGS} only spins on one address. We used a 
buffer size 8 by default but smaller buffers also work. 

There do exist benchmarks where useful work is done during spinning 
and thus more than 8 distinct addresses are involved (\eg, 
\texttt{RADIOSITY}).  Here, livelock detection is 
ineffective and correctness is guaranteed by the self incrementing 
program timestamp.

\vspace{-.05in}
\subsubsection{Livelock Threshold Counter}

\cref{fig:thresh} shows the performance and network traffic normalized 
to the baseline Tardis when sweeping the minimal threshold counter 
(\textit{min\_counter} in Algorithm~\ref{alg:threshold-counter}) in 
the livelock detector. The maximal threshold counter is always 8 times 
of the minimal value. Whenever an address in the AHB has been accessed 
\textit{thresh\_count} times, a check request is sent. With a 
larger \textit{thresh\_count}, checks are sent after spinning for a 
longer time which may hurt performance. On the other hand, larger 
\textit{thresh\_count} can reduce the total number of check messages 
and network traffic. In practice, the \textit{thresh\_count} should be 
chosen to balance the tradeoff. We chose 100 as the default threshold 
counter.  
 
\vspace{-.05in}
\subsubsection{Scalability}

Finally, \cref{fig:256} shows the performance and network traffic 
(normalized to baseline directory) of all benchmarks running on a 
256-core system.  
Compared to the baseline directory, the optimized Tardis 
outperforms by 4.7\% (upto 36.8\%) and reduces the network traffic by 
2.6\% (upto 14.6\%). Although not shown in the graph, we also 
evaluated Tardis and directory where both schemes consume the same 
area overhead. Since Tardis requires less area than directory for 
coherence meta data, it can have a 37\% larger LLC. In area normalized 
evaluation, Tardis can outperform the baseline directory by 6\% on 
average. 

Note that at 256 cores, the performance improvement of Tardis is 
greater than the 64 core case.  This indicates that Tardis not only 
has better scalability in terms of storage as core count increases, it 
also scales better in terms of performance.  

\section{Related Work} \label{sec:related}


Memory coherence is an important issue in shared memory systems with 
private storage in each core or processor. It has been widely studied 
and implemented in multicore processors~\cite{tilera, xeonphi}, 
multi-socket systems~\cite{ziakas2010, anderson2003} and distributed 
shared memory systems~\cite{li1989, keleher1994}.  Traditional
directory or snoopy based coherence protocols enforce the global 
memory order in a consistency model using physical time order. They 
need an invalidation mechanism to guarantee correctness, and therefore 
either require non-scalable storage overhead (e.g., directory-based 
protocols) or broadcasting in the network (e.g., snoopy-based 
protocols). 

Numerous previous works have tried to improve the scalability of 
directory coherence protocols~\cite{agarwal1998, ATAC, chaiken1990, 
maa1991, gupta1990, kelm2010, sanchez2012}. Most of these 
works focused on better ways to organize the directory structure to 
improve scalability. Compared to the full-map directory protocol, 
these optimizations usually hurt performance and increase the design 
and verification complexity.



Several previous works have proposed to simplify the hardware 
coherence protocal for relaxed consistency models like release 
consistency (\cite{choi2011, ros2012, sung2013}) or TSO 
(\cite{elver2014}). Like Tardis, these protocols also do not require 
the sharer list in the directory. Unlike Tardis, they are based on 
the self-invalidation mechanism where all shared cachelines in an L1 
cache need to be invalidated if the consistency model might be 
violated. For release consistency self invalidation happens for every 
synchronization instruction and for TSO it happens for every L1 cache 
miss.

Among papers along this line of research, TSO-CC~\cite{elver2014} is 
the one most similar to Tardis TSO. However, since TSO-CC is 
self-invalidation based, it would incur more L1 misses than Tardis for 
benchmarks with a lot of fences. Another advantage of Tardis over 
TSO-CC is that Tardis is able to efficiently support any consistency 
model with minimal hardware reconfiguration. Running a legacy SC 
program on TSO-CC hardware, however, will end up with suboptimal 
performance and efficiency.  

%
%

\section{Conclusion} \label{sec:conclusion}

In this paper, several optimization techniques have been applied to 
Tardis, a very scalable physiological time based cache coherence 
protocol. Further, more relaxed consistency models such as the Total Store Order (TSO), Partial Store Order (PSO) and Release Consistency (RC) models are now supported in Tardis.
On our set of benchmarks, evaluations indicate that optimized Tardis is better than a full-map directory protocol in terms of performance, energy and storage while being simpler.

{\bf Acknowledgement}: This research was partially supported by MIT PRIMES. We thank Larry Rudolph for helping us improve this paper.


\newpage
\bibliographystyle{ieeetr}
\bibliography{paper}
\footnotesize

\end{document}